\documentclass{PoS}

\title{"Spoon-feeding" an AGN}

\ShortTitle{"Spoon-feeding" an AGN}
\author{\speaker{Deborah Mainetti}\\
	INAF, Osservatorio Astronomico di Brera, Merate (Italy);\\
	Dipartimento di Fisica G.Occhialini, Universit\`{a} degli Studi di Milano Bicocca, Milano (Italy)\\
	E-mail: \email{deborah.mainetti@brera.inaf.it}}

\author{Sergio Campana\\
	INAF, Osservatorio Astronomico di Brera, Merate (Italy)\\
	Email: \email{sergio.campana@brera.inaf.it}}
	
\author{Monica Colpi\\
	Dipartimento di Fisica G.Occhialini, Universit\`{a} degli Studi di Milano Bicocca, Milano (Italy)\\
	E-mail: \email{Monica.Colpi@mib.infn.it}}
	
\author{Giuseppe Lodato\\
	Dipartimento di Fisica, Universit\`{a} degli Studi di Milano, Milano (Italy)\\
	E-mail: \email{giuseppe.lodato@unimi.it}}
	
\author{Paolo D'Avanzo\\
	INAF, Osservatorio Astronomico di Brera, Merate (Italy)\\
	E-mail: \email{paolo.davanzo@brera.inaf.it}}
	
\author{Philip Andrew Evans\\
	Department of Physics and Astronomy, University of Leicester (UK)\\
	E-mail: \email{pae9@leicester.ac.uk}}
	
\author{Alberto Moretti\\
	INAF, Osservatorio Astronomico di Brera, Milano (Italy)\\
	E-mail: \email{alberto.moretti@brera.inaf.it}}

\abstract{Tidal disruption events (TDEs) occur when a star, passing too close to a massive black hole, is ripped apart by tidal forces. A less dramatic event occurs if the star orbits just outside the tidal radius, resulting in a mild stripping of mass. Thus, if a star orbits a central black hole on one of these bound eccentric orbits, weaker outbursts will occur recurring every orbital period. Thanks to five Swift observations, we observed a recent flare from the close by (92 Mpc) galaxy IC 3599, where a possible TDE was already observed in December 1990 during the Rosat All-Sky Survey. By light curve modeling and spectral fitting, we account for all these events as the non-disruptive tidal stripping of a single star into a 9.5 yr highly eccentric bound orbit. This is the first example of periodic partial tidal disruptions, possibly spoon-feeding the central black hole.}

\FullConference{Swift: 10 Years of Discovery,\\
		2-5 December 2014\\
		La Sapienza University, Rome, Italy }

\begin{document}

%\section{...}
\section{Introduction: total and partial tidal disruption events and periodic partial stripping of stars}
Tidal disruption events occur when a star passes close to a massive black hole and it is totally disrupted \cite{Rees},\cite{Phinney} or partially stripped \cite{MacLeod} by tidal forces. The distinction between total and partial tidal events is based on the relation between the pericenter radius of the star ($R_{p}$) and the so-called tidal radius of the black hole, i.e. on the impact parameter $\beta = \frac{R_{t}}{R_{p}}$ \cite{Guillochon}. The tidal radius is the radius at which the star self-gravity equates the black hole tidal force and it is defined as $R_{t} \sim R_{*}(\frac{M_{BH}}{M_{*}})^{\frac{1}{3}}$, where $R_{*}$ and $M_{*}$ are the star radius and mass and $M_{BH}$ is the black hole mass.

Let's start considering a star on a parabolic orbit around a massive black hole. If its pericenter radius is less than about the tidal radius of the black hole (i.e. $\beta \gtrsim 1$), the black hole tidal force overcomes the star self-gravity and the star is totally destroyed. About half of the stellar debris leaves the system on hyperbolic orbits, while the other half remains bound to the system and accretes onto the black hole through an accretion disk. As a consequence of accretion, a bright flare is produced. According to the classical theory of tidal disruption events, the trend of the flare bolometric luminosity in time is described by a power law of index $-\frac{5}{3}$ \cite{Rees}, \cite{Phinney}, \cite{Lodato}.

If the star pericenter radius is greater than about the black hole tidal radius ($\beta \lesssim 1$) a less dramatic event occurs. The star is not totally disrupted but only a fraction of its mass is stripped by the black hole. The dynamics of partial tidal events are quite the same of total tidal events.

A noteworthy case is the one involving a star on an eccentric and bound orbit around a massive black hole in the regime of partial tidal events. The star could transfer a fraction of its mass to the black hole every time it passes through its pericenter. In this way a flare for each orbital period could be observed. Over several orbital periods this mass transfer might also feed the quiescent luminosity of low-luminosity active galaxies. Such a feeding has been theorised and named "spoon-feeding" in the literature \cite{MacLeod}, but no periodic partial disruption event has ever been observed yet. Periodic partial events should be more frequent than classical tidal disruption events (a factor of $\sim 10$) and they should repeat over the orbital period of the star, but they are less energetic than classical tidal events because of the lower level of mass transfer. 

\section{The case of IC 3599}
A candidate periodic sequence of partial disruption events has been observed in the spiral galaxy IC 3599 (see \cite{Campana} for more details). This is a weakly active galaxy in the Coma Cluster ($\sim 92$ Mpc). IC 3599 was discovered as a soft X-ray source during the ROSAT All Sky Survey in December 1990. Following ROSAT and Chandra observations from 1992 to 2002 showed an X-ray flux decrease of a factor of about 100 and an X-ray spectral hardening. Several authors \cite{Grupe}, \cite{Brandt}, \cite{Komossa}, \cite{Vaughan} interpreted this emission as coming from the total tidal disruption of a star orbiting around the central massive black hole ($\sim 3 \times 10^5 M_{\odot}$ \cite{Sani}) of the galaxy, even if IC 3599 is a low-luminosity active galactic nucleus (AGN). Indeed in AGNs it is difficult to distinguish between variability coming from the galaxy itself and variability induced by the disruption of a star, even if in AGNs the presence of a pre-existent accretion disk enhances the probability to have tidal disruption events \cite{Komossa1}, \cite{Karas}.

From 2010 to 2014 IC 3599 was observed again by the Swift/XRT telescope. Surprisingly this source showed a new increase and successive decrease of a factor of about 30 in its bolometric luminosity an also a new softening and then hardening in its X-ray spectrum very similarly to the already observed event. We tried to evaluate which could be the cause of such an emission.

The AGN nature of IC 3599 may raise doubts about the tidal disruption interpretation because like other AGNs it could emit flares due to (slim) disk instabilities or due to the uncovering of a heavily absorbed X-ray source \cite{Grupe1}. The first hypothesis was excluded because the estimated mass accretion rate for IC 3599 (see below) is at the very border of the instability region \cite{Honma} and the duration of the flares is too long for theory predictions \cite{Xue}. The second hypothesis was excluded because of  the non-acceptable spectral fit of ROSAT, Chandra and Swift/XRT observations with a model which includes a Galactic absorption plus a partial covering of a power law component ($\chi^2=627.4$, 304 d.o.f.). We also ruled out AGN variability to explain the observed emission in IC 3599. Firstly we took the power spectral density of a typical AGN with a black hole of mass $\sim 3 \times 10^5 M_{\odot}$ (similar to the one in IC 3599) \cite{Mchardy}, \cite{Mchardy1} and from this we simulated light curves also including an unrealistic $100\%$ root mean squared variability. We found a maximum variability of  $\sim 6-7$ which is too small. Secondly we considered a group of AGNs selected in the literature \cite{Grupe2}, \cite{Grupe3}, \cite{Grupe4} and a group monitored by Swift/XRT with a variability of at least a factor of 20 and from each of their light curves we randomly extracted 7 and 6 points for 500,000 times, fitting them separately with a power law of index $-\frac{5}{3}$ plus a constant in order to simulate the first and the second luminosity decays observed for IC 3599 (see below). We counted the number of simulations with a null hypothesis probability larger than $5\%$ and based on this number we evaluated the probability of obtaining by chance a TDE-like light curve. Then we evaluated the probability to have the best obtained $\chi^2$ with respect to the number of degrees of freedom, weighting it for the number of trials. In both these procedures we obtained that a variability similar to the one observed in IC 3599 occurs by chance at $\sim 4.5 \sigma$ level. We could safely conclude that the observed variability in IC 3599 do not come from known AGN variability. Again we excluded the total (or partial not periodic) disruption of a single star and also the disruption of multiple stars, because in the first case we should have observed only one emission peak while in the second case we should have likely observed different light curves. Then we excluded a binary made of a black hole and a companion star because an accretion through wind from the companion star would be too inefficient to produce the observed luminosity while an accretion through Roche lobe overflow would imply the companion star to be too underdense \cite{Lasota}. We also excluded a binary black hole because in this case, according to the mass of the central black hole in IC 3599, we would not predict emission in the X-ray band \cite{Tanaka}, at variance with observations. A number of other models for IC 3599, either AGN-disk related, or TDE-related, or binary super-massive BH-related were explored by Komossa et al. (these proceedings).

\begin{figure}[h!]
%\centering
\includegraphics[width=160mm,angle=0,clip=]{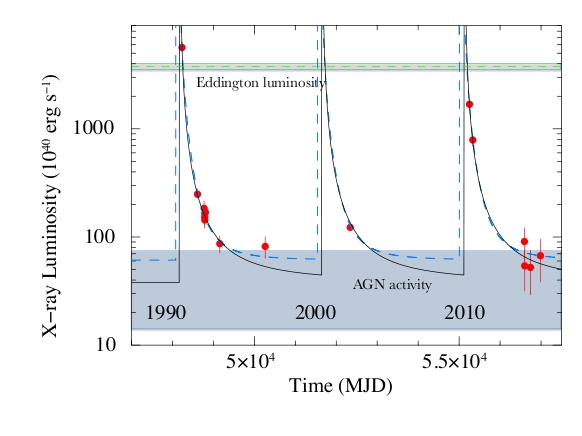}
\caption{Bolometric luminosity of IC 3599 from ROSAT, Chandra and Swift/XRT observations from 1990 to 2014. The fit with solid line is the case of three equal power laws of index $-\frac{5}{3}$, while the fit with dashed line is the case of three equal power laws with free index. \label{luminosity}}
\end{figure}

The hypothesis of periodic partial tidal events seems to be in agreement with the observations. We fitted ROSAT, Chandra and Swift/XRT observations with single component models and also with two double component models, one including a Galactic absorption and a partial covering of a power law component (as discussed above) and the other including an accretion disk component and a power law component. The last model describes a tidal disruption event (the disk component stays for the emission caused by the tidal debris accretion, while the power law component is for the quiescent AGN activity) and is also the statistically best one ($\chi^2=315.9$, 303 d.o.f.). We fitted the bolometric light curve of IC 3599 with a series of equal power laws of index $-\frac{5}{3}$ and we obtained the statistically best fit in the case of three equal power laws ($\chi_{red}^2 = 0.83$, 11 d.o.f.), with an orbital period coming from the fit of about 9.5 years (Figure \ref{luminosity}). 
Then we let free the index of the power law and we obtained a range of values that is fully in agreement with the $-\frac{5}{3}$ value ($-2.7 \pm 1.1$, $90\%$ confidence level; Figure \ref{luminosity}). A fit with an exponential function for the first and the third emission peaks provides a worse fit but highlights the similarities between them, being characterised by a decay time of $107 \pm 7$ d and $100 \pm 42$ d respectively ($90\%$ confidence level). This fit also corroborates the idea that the observed emission peaks are related and do not come from AGN variability. 

\begin{figure}[h!]
\centering
\includegraphics[width=120mm,angle=0,clip=]{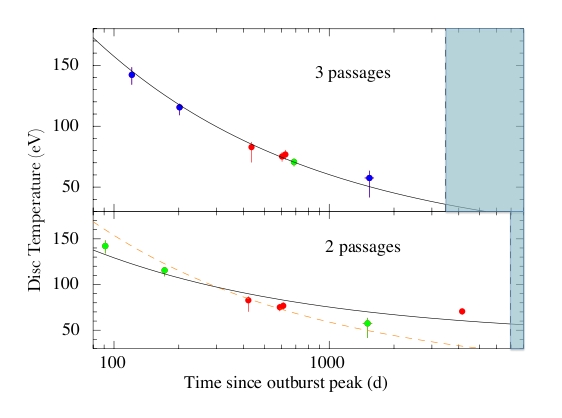}
\caption{Folding of the disk temperatures, obtained from spectral fit. The upper panel is the case of three emission peaks (red points for the first passage, green point for the second passage, blue points for the third passage), while the bottom panel is the case of two emission peaks (red points for the first passage, green points for the second passage). The fit with orange line in the bottom panel is without a constant. The fit in the upper panel is without a constant.  \label{temperature}}
\end{figure}

We also folded over the orbital period the disk temperatures of the considered source derived from the spectral fit, in the case of a light curve fitted with two and three equal power laws, and we fitted these trends with a power law of index $-\frac{5}{12}$ (coming from $\dot M \propto t^{-\frac{5}{3}}$ and $T \propto \dot M^{\frac{1}{4}}$, typical of an accretion disk). We obtained the statistically best fit in the case of three equal power laws again ($\chi_{red}^2=0.6$, 6 d.o.f.; Figure \ref{temperature}).

\begin{figure}[h!]
\centering
\includegraphics[width=150mm,angle=0,clip=]{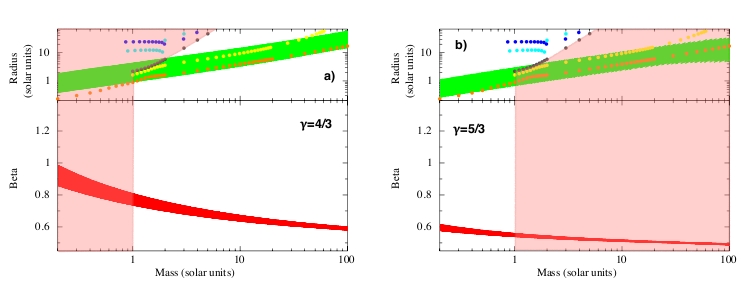}
\caption{Possible combinations of mass, radius and impact parameter for the involved star. Light red regions are the not considered ones. The left panel is the case of a polytope of index $\frac{4}{3}$ (massive main sequence stars or stars just coming out the main sequence phase), the right panel is the case of a polytope of index $\frac{5}{3}$ (low-mass main sequence stars or giant stars). Coloured points are for different evolutionary phases of stars with different initial masses (coming out from the use of the SSE code): orange points stay for main sequence stars, yellow points for stars on the Hertzsprung gap, black points for giant stars. \label{star}}
\end{figure}

Finally, we investigated the possible nature of the star involved in the tidal stripping through fitting formulae coming from hydrodynamical simulations already present in the literature \cite{Guillochon}, \cite{MacLeod1}, using our estimation of the fallback rate during the first observation of the source ($\dot M_{peak} \sim 0.01 M_{\odot} $ $\rm yr^{-1}$) and of the accreted mass per episode ($\sim 2.5 \times 10^{-3} M_{\odot}$) and approximating the star as a polytope. In this way we could obtain the possible combinations of mass, radius and impact parameter that characterise the star. We also simulated the evolution of stars with different initial masses  by means of the Single Stellar Evolution code \cite{Hurley}. In the case of a polytope of polytropic index $\frac{4}{3}$ (massive stars on the main sequence phase or just coming out the main sequence) we found values for the impact parameter below 1.85 (0.60-0.81), while in the case of a polytope of polytropic index $\frac{5}{3}$  (low-mass stars on the main sequence phase or giant stars) we found values of $\beta$ below 0.9 (0.54-0.58) (Figure \ref{star}). 
These are the critical values above which only total tidal events occur and below which tidal events are partial \cite{Guillochon}. We considered a star of $M_{*}=10 M_{\odot}$ ($R_{*}=7.5 R_{\odot}$, $\beta=0.65$), in analogy with the S-stars around the centre of our Galaxy, and we evaluated its eccentricity. We found a very high value ($\sim 0.995$), but not so strange considering the predicted distribution of eccentricities around the centre of the Milky Way \cite{Gillessen}.

So we are confident to have found the first observed periodic sequence of partial tidal disruption events, maybe also with the "spoon-feed" of the AGN \cite{MacLeod}. According to our calculations, the involved star should be totally consumed in about 10,000 years and the next emission peak should be observable in 2019.

\end{document}